\documentstyle[twocolumn,floats,aps]{revtex}
\input psfig
\begin{document}
\draft

\title{Pinning Control of Spatiotemporal Chaos}
\author{R. O. Grigoriev, M. C. Cross and H. G. Schuster}
\address{Condensed Matter Physics 114-36
and Neural Systems Program 139-74\\
California Institute of Technology, Pasadena CA 91125 }
\date{\today}
\maketitle

\begin{abstract}
 Linear control theory is used to develop an improved localized control
scheme for spatially extended chaotic systems, which is applied to a
Coupled Map Lattice as an example. The optimal arrangement of the control
sites is shown to depend on the symmetry properties of the system, while
their minimal density depends on the strength of noise in the system. The
method is shown to work in any region of parameter space and requires a
significantly smaller number of controllers compared to the method
proposed earlier by Qu and Hu \cite{gang}. A nonlinear generalization of
the method for a 1-d lattice is also presented.
 \end{abstract}


\narrowtext

Controlling chaos in high-dimensional systems \cite{ding} and
spatiotemporal chaos especially is a very important problem with numerous
applications to turbulence \cite{katz}, instabilities in plasma
\cite{pentek}, multi-mode lasers \cite{colet} and reaction-diffusion
systems \cite{crowley}.

The present letter represents an effort to develop a general control
algorithm for spatiotemporally chaotic systems using the methodology of
linear control theory, which already proved to be fruitful
\cite{romeiras}. Clarifying a number of issues will have direct bearing
on this. For instance, it is not clear how many parameters are required
for successful control. If the control is applied locally, what is the
minimal density of controllers and how should they be arranged to obtain
optimal performance?  What are the limitations of the linear control
scheme and how can they be overcome? 

Consider the Coupled Map Lattice (CML), originally introduced by Kaneko
\cite{kaneko}:
 \begin{eqnarray}
 \label{eq_cml}
 z_i^{t+1}&=&F(z_{i-1}^t,z_i^t,z_{i+1}^t)\cr
 &=&f((1-2\epsilon)z_i^t+\epsilon(z_{i-1}^t+z_{i+1}^t)),
 \end{eqnarray}
 with $i=1,2,\cdots,L$ and periodic boundary conditions, i.e. 
$z_{i+L}^t=z_i^t$ imposed. We also assume that the local map $f(z,a)$ is a
nonlinear function with parameter $a$, such that $f(z^*,a^*)=z^*$.

To be specific, we choose
 \begin{equation}
 \label{eq_loc_map}
 f(z)=az(1-z),
 \end{equation}
 but emphasize, that all the major results hold independent of this
choice. This CML has a homogeneous steady state $z^*=1-{1 \over a}$, which
is unstable for $a>3.0$ and our goal is to stabilize it using a minimal
number of controllers. 

The first attempt in this direction was undertaken by Hu and Qu
\cite{gang}. The authors tried to stabilize the homogeneous state by
controlling an array of $M$ {\sl periodically} placed pinning sites
$\{i_1,\cdots,i_M\}$ with appropriately chosen control $u_m^t$
 \begin{equation}
 \label{eq_cml_ctrl}
 z_i^{t+1}=F(z_{i-1}^t,z_i^t,z_{i+1}^t)+\sum_{m=1}^M\delta(i-i_m)u_m^t. 
 \end{equation}
This however required a very dense array with distance between controllers
$L_p=L/M\le 3$ in the physically interesting interval of parameters
$3.57<a<4.0$. 

The reason for this is the spatial periodicity of the pinnings. Since the
system is spatially uniform, its eigenmodes are just Fourier modes and the
pinning sites do not affect the modes whose nodes happen to lie at the
pinnings, i.e. modes with periods equal to $2L_p$, $2L_p/2$, $2L_p/3$,
etc, provided those are integer. The control scheme worked only when {\sl
all} such modes were {\sl stable}.

It is however not necessary to destroy the periodicity completely to
achieve control: that would complicate the analysis unnecessarily. Instead
we {\sl add} one more pinning site between each of the existing ones. Not
all positions are good, but some do solve the problem --- previously
uncontrollable modes become controllable. 

In order to understand how the pinnings should be placed and see whether
we achieve improved performance by introducing additional controllers, we
have to use a few results of the linear control theory \cite{dorato}. We
will start with linearizing eq. (\ref{eq_cml_ctrl}) about the homogeneous
steady state ${\bf z}^t=(z^*,\cdots,z^*)$ in both the state vector and
control to obtain the following standard equation
 \begin{equation}
 \label{eq_gen_lin}
 {\bf x}^{t+1}=A{\bf x}^t+B{\bf u}^t,
 \end{equation}
 where we denoted the displacement ${\bf x}={\bf z}-{\bf z}^*$. If we
define $\alpha=\partial f(x^*,a^*) / \partial x$, then the $L\times L$
Jacobian $A$ is given by
 \begin{equation}
 \label{eq_matr_a}
  A = \alpha \left( \matrix{
      1-2\epsilon & \epsilon & 0 & \cdots &\epsilon \cr
      \epsilon & 1-2\epsilon & \epsilon & \cdots & 0 \cr
      0 & \epsilon & 1-2\epsilon & \cdots & 0 \cr
      \vdots & \vdots & \vdots & \ddots & \vdots\cr
      \epsilon & 0 & 0 & \cdots & 1-2\epsilon \cr
    } \right)
 \end{equation}
 and the $L\times M$ control matrix $B_{ij}=\delta(j-m)\delta(i-i_m)$
depends on how we place the pinning sites.

If we use synchronous linear feedback ${\bf u}^t=-K{\bf x}^t$, equation
(\ref{eq_gen_lin}) becomes
 \begin{equation}
 {\bf x}^{t+1}=(A-BK){\bf x}^t,
 \end{equation}
 and the solution ${\bf x}=0$ can be made stable by a suitable choice of
the feedback gain matrix $K$, if the {\sl controllability} condition ${\rm
rank}(C)=L$ is satisfied. The controllability matrix $C$ is defined via
 \begin{equation}
 \label{eq_ctblm}
 C=(B\ AB\ \cdots\ A^{L-1}B).
 \end{equation}

 One can easily verify that the matrix $B$ calculated for a periodic array
of pinning sites does not satisfy the controllability condition and
therefore the homogeneous steady state is not controllable. It can be
stabilized if the weaker {\sl stabilizability} condition is satisfied,
i.e. all uncontrollable modes are stable. However this imposes excessive
restrictions on the pinning density. 

\begin{figure} 
\psfig{figure=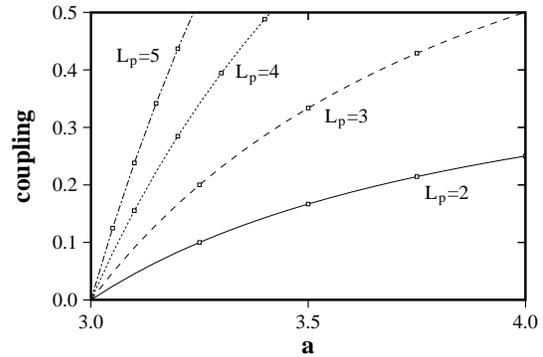,width=3.5in}
 \caption{ Periodic array of single pinning sites: critical coupling
$\epsilon_{cr}$ as a function of parameter $a$. The dots represent the
numerical results from figure 2 of [1], with $\epsilon$ rescaled by a
factor of two to make it compatible with our definition. }
 \label{fig_coupling}
\end{figure}

The condition for stabilizability can be obtained from the spectrum of 
eigenvalues of the matrix (\ref{eq_matr_a})
 \begin{equation}
 \label{eq_spectrum}
 \gamma_i=\alpha(1-2\epsilon(1-\cos(k_i)),
 \end{equation}
 where $k_1=0$, $k_i=k_{i+1}=\pi i/L$ for $i=2,4,6,\cdots$ and, for
$L$-even, $K_L=\pi$ and $\alpha=2-a$. Specifically, we need
$|(a-2)(1-2\epsilon(1-\cos(\pi j/L_p))|<1$ for all $j=1,\cdots,L-2$, such
that $L_p/j$ is integer. Using this criterion one can obtain the relation
between the minimum coupling, the distance between controllers and
parameter $a$ of the local chaotic map for a stabilizable system. For
instance, $j=1$ yields
 \begin{equation}
 \label{eq_min_coup}
 \epsilon={a-3 \over 2(a-2)(1-\cos(\pi/L_p))}.
 \end{equation}
 The results are presented in fig. \ref{fig_coupling}. It can be easily
verified that they coincide with the numerically obtained results of
Hu and Qu.

It is possible however to extend the limits of the control scheme quite
substantially by making the system {\sl controllable} as opposed to {\sl
stabilizable}. This is easily achieved by choosing a different matrix $B$,
i.e. placing the pinning sites differently. Doing so will enable us to
control the system {\sl anywhere} in the parameter space at the same time
using a {\sl smaller} density of controllers. 

First one has to determine the dimensionality of the matrix $B$, in other
words determine the minimal number of parameters required to control the
CML (\ref{eq_cml}) of an arbitrary length. It can be shown \cite{self}
that the minimal number of parameters required to control a system with
degenerate Jacobian is equal to the greatest multiplicity of its
eigenvalues.

Since the system under consideration has parity symmetry, the eigenvalues
(\ref{eq_spectrum}) of its Jacobian are in fact doubly degenerate, so the
minimal number of control parameters yielding a controllable system in our
case is two, meaning at least two pinning sites are required. One can
easily verify that the controllability condition for an $L\times 2$ matrix
 \begin{equation}
 B_{ij}=\delta(j-1)\delta(i-i_1)+\delta(j-2)\delta(i-i_2)
 \end{equation}
 is indeed satisfied for a number of arrangements $\{i_1,i_2\}$. The
restrictions on the mutual arrangement of the controllers are again given
by the condition of controllability: $L$ should not be a multiple of
$|i_2-i_1|$, otherwise the mode with the period $2|i_2-i_1|$ becomes
uncontrollable. 

The next step in the algorithm is to determine the feedback gain $K$. Pole
placement techniques based on Ackermann's method \cite{barreto} are
inapplicable to the problem of controlling spatially extended systems
because they are {\sl numerically unstable} \cite{kautsky} and break down
rapidly for problems of order greater than 10.

Instead we use the method of the linear-quadratic (LQ) control theory
\cite{dorato}, applicable to the unstable periodic trajectories as well as
fixed points. This method is not only numerically stable, but also allows
one to {\sl optimize} the control algorithm to increase convergence speed,
and at the same time minimize the strength of control. As we will see
below, decreasing control enlarges the basin of attraction, which has very
important consequences for the time to achieve control (capture the
chaotic trajectory). The optimal solution is obtained by minimizing the
cost functional
 \begin{equation}
 \label{eq_functional}
 V({\bf x}^0)=\sum_{n=0}^{\infty}
 ({{\bf x}^t}^{\dagger}Q{\bf x}^t+{{\bf u}^t}^{\dagger}R{\bf u}^t),
 \end{equation}
 where $Q$ and $R$ are the weight matrices that can be chosen as any
positive-definite square matrices.

The minimum of (\ref{eq_functional}) is reached when
 \begin{equation}
 \label{eq_feedback}
 K=(R+B^{\dagger}PB)^{-1}B^{\dagger}PA,
 \end{equation}
where $P$ is the solution to the discrete-time algebraic Ricatti equation
 \begin{equation}
 \label{eq_ricatti}
 P=(Q+A^{\dagger}PA)-A^{\dagger}P^{\dagger}B(R+B^{\dagger}PB)^{-1}
 B^{\dagger}PA.
 \end{equation}

\begin{figure} 
\psfig{figure=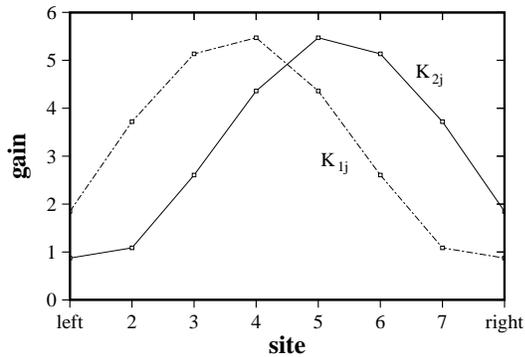,width=3.5in}
 \caption{ Feedback matrix: feedback gain vectors $K_{1j}$ and $K_{2j}$
for left and right controller placed at the sides of the lattice ($i_1=1,
i_2=8$) as functions of the lattice site for $a=4.0$ and $\epsilon=0.33$. }
 \label{fig_feedback}
\end{figure}

Numerical simulations show that the CML (\ref{eq_cml},\ref{eq_loc_map}) 
can indeed by stabilized by this linear control scheme in a wide range of
parameters $a$ and $\epsilon$. The solution for $K$ is presented in figure
\ref{fig_feedback} for $a=4.0$, $\epsilon=0.33$ and $L=8$ with
$Q=I_{8\times 8}$ and $R=I_{2\times 2}$. The steady homogeneous state
$z^*=0.75$ has 3 unstable and 5 stable directions and we use 2 pinning
sites to control it. 

The contribution $-K_{mi}x^n_i$ from the sites $i$ far away from the
pinning site $i_m$ is larger, as one would expect: since the feedback is
applied indirectly through coupling to the neighbors, the perturbation
introduced by the controllers decays with increasing distance to the
pinning sites. 

Noise limits our ability to locally control arbitrarily large systems with
local interactions. We will use a simple illustrative approach to see the
effect of noise on the control scheme. The rank of the matrix is given by
the number of its nonzero singular values. The singular values of the
controllability matrix (\ref{eq_ctblm}) scale roughly as $s_l\sim
|\gamma_1|^l$, where $\gamma_1$ is the largest eigenvalue of the Jacobian
$A$
 \begin{equation}
 |\gamma_1|=e^{\lambda_{max}}=\cases{
      |\alpha|, & $\epsilon<0.5$,\cr
      |\alpha(4\epsilon-1)|, & $\epsilon>0.5$.}
 \end{equation} 
 Assuming that there is an uncertainty in the calculations (due to the
uncertainty in the state vector, parameter vector or just numerical
roundoff errors) of relative magnitude $\sigma$, we can say that the rank
of the controllability matrix can be reliably determined to be equal to
the length of the lattice if $s_0/s_L>\sigma$. This gives us the
theoretical bounds on the size of the controllable system in the presence
of noise: 
 \begin{equation}
 \label{eq_rank_bnd} 
 L_{max}^{(1)}=-{\log(\sigma)\over\lambda_{max}}.
 \end{equation} 

On the other hand, the perturbation $\delta x_i$ introduced by the
controller $i$ affects the dynamics of the remote site $j$ after
propagating a distance $\Delta=|i-j|$ in time $\tau=\Delta$, decaying by a
factor of $\epsilon$ per iteration, while the noise at site $j$ increases
roughly by a factor of $\gamma_1$ per iteration. We therefore need $\delta
x_i\epsilon^\Delta> \sigma|\gamma_1|^\tau$. Since the maximum distance
$\Delta$ to the closest controller is $L/2$ and $\delta x_i\sim 1$, we get
another bound, complementing (\ref{eq_rank_bnd}) 
 \begin{equation}
 \label{eq_coup_bnd}
 L_{max}^{(2)}={2\log(\sigma)\over\log(\epsilon)-\lambda_{max}}.
 \end{equation} 
 Similar constraints were obtained by Aranson et. al for the lattices with
asymmetric coupling (cf. equation (15) of the ref. \cite{aranson}).

\begin{figure}
\psfig{figure=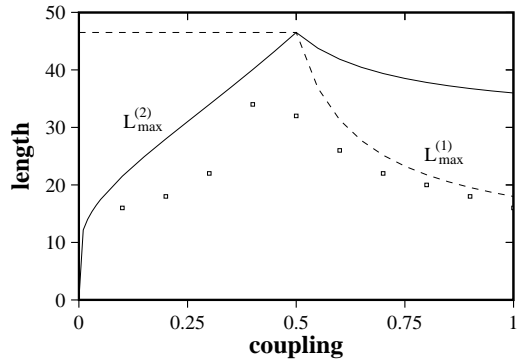,width=3.5in}
 \caption{ The largest length of the lattice which can be controlled with
two pinning sites:  theoretical estimates (solid lines) and numerical
results (dots) obtained with the uniformly distributed noise of amplitude
$\sigma=10^{-14}$ as functions of coupling $\epsilon$ for $a=4.0$. }
 \label{fig_length}
\end{figure}

The maximal length of the system, that can be stabilized by the LQ method
with two pinning sites placed next to each other is obtained numerically
by choosing the initial condition very close to the fixed point ($|{\bf
x}^0|\ll|\gamma_1|^{-L/2}$) and letting the system evolve under control
(\ref{eq_feedback}) calculated for $Q=I_{L\times L}$ and $R=I_{2\times
2}$. This length is quite large even in the presence of noise (fig. 
\ref{fig_length}) and agrees with the theoretical bounds
(\ref{eq_rank_bnd}, \ref{eq_coup_bnd}) rather well for such a crude
estimate. 

The problem of controlling a large 1-dimensional system with the length
$L>L_{max}(\sigma)$ exceeding the maximum allowed for a given noise level
can be easily reduced to the problem of controlling a number of smaller
systems with the length $L_p<L_{max}(\sigma)$. We partition the entire
lattice $\{z^t_1,\cdots,z^t_L\}$ into $M=L/L_p$ subdomains
$\{z^t_{(m-1)L_p+1}, \cdots,z^t_{mL_p}\}$, and control it with an array of
pinning sites $i_{m1}=(m-1)L_p+1$, $i_{m2}=mL_p$, $m=1,\cdots\,M$
positioned periodically at the boundaries of these subdomains.

The stabilization can be achieved by choosing
 \begin{eqnarray}
 \label{eq_nonlin}
 &u^t_{i_{m1}}&=F(z_{i_{m2}}^t,z_{i_{m1}}^t,z_{i_{m1}+1}^t)-
 F(z_{i_{m1}-1}^t,z_{i_{m1}}^t,z_{i_{m1}+1}^t)\cr
 &+&\prod_{i=1}^{L_p}\theta(\delta x_i-|x^t_{(m-1)L_p+i}|)
 \sum_{i=1}^{L_p}K_{1i}x^t_{(m-1)L_p+i}\cr
 &u^t_{i_{m2}}&=F(z_{i_{m2}-1}^t,z_{i_{m2}}^t,z_{i_{m1}}^t)-
 F(z_{i_{m2}-1}^t,z_{i_{m2}}^t,z_{i_{m2}+1}^t)\cr
 &+&\prod_{i=1}^{L_p}\theta(\delta x_i-|x^t_{(m-1)L_p+i}|)
 \sum_{i=1}^{L_p}K_{2i}x^t_{(m-1)L_p+i},
 \end{eqnarray}
 where $\theta(x)$ is a step-function.

This arrangement effectively carries two functions. We use control
(\ref{eq_nonlin}) to (nonlinearly) decouple the subdomains, simultaneously
imposing periodic boundary condition for each subdomain (the first two
terms) to make the system controllable. Then we stabilize each subdomain
asynchronously by applying a linear (in deviation $x_i^t=z_i^t-z^*$) 
feedback (the last term), inside the neighborhood of the fixed point
determined by $\delta x_i$. The linear approximation (\ref{eq_gen_lin})
is only valid if
 \begin{equation}
 \delta x_i\ll |K_{mi}|^{-1},\qquad m=1,\cdots,M 
 \end{equation}
 and therefore strong feedback significantly decreases the size of the
capture region, which makes the capture time vary large. Minimizing the
capture time can be achieved by minimizing the feedback strength using the
LQ method (\ref{eq_feedback},\ref{eq_ricatti}).

\begin{figure} 
\psfig{figure=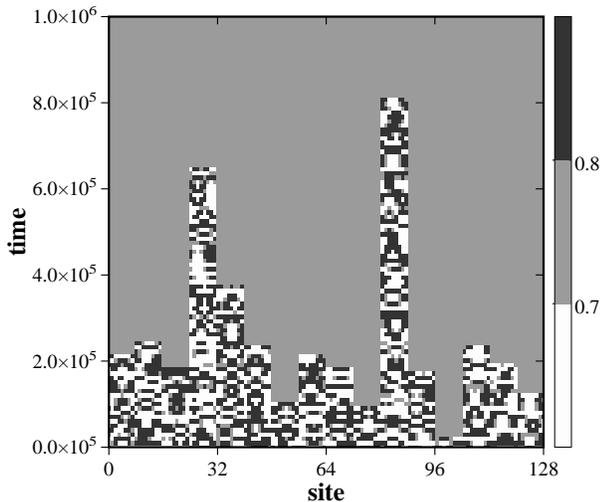,width=3.5in}
 \caption{ Stabilizing uniform steady state: a large lattice ($L=128$) is
controlled by an array of double pinning sites, placed at the boundaries
of subdomains with length $L_p=8$. The state of the system was plotted at
each 10000-th step. }
 \label{fig_multiple}
\end{figure}

We demonstrate this approach by stabilizing the homogeneous stationary
state of the CML defined by equation (\ref{eq_cml},\ref{eq_loc_map}) with
$a=4.0$, $\epsilon=0.33$.  $L=128$ sites were divided into $M=16$
subdomains of length $L_p=8$, each controlled by two pinning sites. The
results presented in fig. \ref{fig_multiple} show the evolution of the
system from the initial condition chosen to be a collection of random
numbers in the interval $[0,1]$. 

Eqs. (\ref{eq_rank_bnd},\ref{eq_coup_bnd}) now give the minimal density of
pinning sites that yields the controllable fixed point solution. It is
indeed seen to be much lower than that given by (\ref{eq_min_coup}), e.g.
$2/L_p=1/20$ ($1/17$ from the numerics, see fig. \ref{fig_length}) as
opposed to $1/L_p=1/2$ for the choice $a=4.0$, $\epsilon=0.4$ and the
precision of calculations given by $\sigma=10^{-14}$.

Although the resulting control scheme becomes nonlinear (and therefore
requires full knowledge of the evolution equations), it has the additional
benefit, that the capture time is determined by the length $L_p\ll L$ and
is typically many orders of magnitude smaller than that obtained for the
linear control scheme (obtained by linearizing (\ref{eq_nonlin})), which
only requires the Jacobian to be known. In fact our computational
resources were insufficient to observe even a single capture for $L>40$
with the linearized control. Generalizing this nonlinear approach to
higher-dimensional systems remains a challenge.

To summarize, we have shown that the restrictions on the minimal density
of periodically placed single pinning sites obtained by Qu and Hu
\cite{gang} as a result of numerical simulations can in fact be obtained
analytically from the stabilizability condition.

The efficiency of the control scheme can be improved significantly if one
uses double pinnings instead of single ones. The homogeneous steady state
becomes controllable for any values of the control parameters and the
minimal density of pinning sites is reduced substantially. It is shown
that the maximal distance between the pinnings depends on the strength of
noise in the system and can be estimated analytically. 

The appropriately chosen (using the LQ technique) feedback can decrease
the capture time for the chaotic trajectory by enlarging the capture
region. The introduction of nonlinearity into the control scheme can
decrease this time even more significantly by effectively decoupling the
large lattice into a number of smaller subdomains. 

The authors thank Prof. J.C. Doyle for many fruitful discussions. This
work was partially supported by the NSF through grant no. DMR-9013984. 
H.G.S. thanks C. Koch for the kind hospitality extended to him at Caltech
and the Volkswagen Foundation for financial support.

\onecolumn

\widetext

\vfill\eject


\begin{references}

\bibitem{gang} G. Hu and Z. Qu,  Phys. Rev. Lett. {\bf 72} (1994) 68. 

\bibitem{ding} M. Ding et al., Phys. Rev. E {\bf 53} (1996) 4334;
 Y. C. Lai and C. Grebogi, Phys. Rev. E {\bf 50} (1994) 1894;
 J. Warncke, M. Bauer and W. Martienssen,  Europhys. Lett. {\bf 25} (1994)
323; D. Auerbach, Phys. Rev. Lett. {\bf 72} (1994) 1184.

\bibitem{katz} R. A. Katz, T. Galib and J. Cembrola, J. Phys. IV {\bf 4}
(1994) 1063.

\bibitem{pentek} A. Pentek, J. B. Kadtke and Z. Toroczkai, Phys. Lett. A
{\bf 224} (1996) 85.

\bibitem{colet} P. Colet, R. Roy and K. Weisenfeld, Phys. Rev. E {\bf 50}
1994) 3453.

\bibitem{crowley} V. Petrov, M. J. Crowley and K. Showalter, Phys. Rev. 
Lett. {\bf 72} (1994) 2955.

\bibitem{romeiras} F. J. Romeiras, et al., Physica D {\bf 58} (1992) 165; 
 V. Petrov et al., Phys. Rev. E {\bf 51} (1995) 3988.

\bibitem{kaneko} K. Kaneko,  Prog. Theor. Phys. {\bf 72} (1984) 480. 

\bibitem{dorato} P. Dorato, C. Abdallah and V. Cerrone, {\sl
Linear-Quadratic Control: An Introduction} (Prentice Hall, New Jersey,
1995) 

\bibitem{self} R. O. Grigoriev and H. G. Schuster (in preparation).

\bibitem{barreto} E. Barreto and C. Grebogi,  Phys. Rev. E {\bf 52} 
(1995) 3553. 

\bibitem{kautsky} J. Kautsky, N. K. Nichols and Van Dooren, Intl. J. 
Control, {\bf 41} (1985), 1129.

\bibitem{aranson} I. Aranson, D. Golomb and H. Sompolinsky,
Phys. Rev. Lett. {\bf 68} (1992) 3495. 

\end{references}
\end{document}